\documentclass[lettersize,journal]{IEEEtran}
\usepackage{amsmath,amssymb,amsfonts,dsfont}
\usepackage[linesnumbered,ruled]{algorithm2e}
\usepackage{array}
\usepackage[caption=false,font=normalsize,labelfont=sf,textfont=sf]{subfig}
\usepackage{textcomp}
\usepackage{stfloats}
\usepackage{url}
\usepackage{verbatim}
\usepackage{graphicx}
\usepackage{cite}
\usepackage{amsthm}
\usepackage{mismath}
\usepackage{caption}
\usepackage{xcolor}
\usepackage{grffile}
\usepackage{paralist}
\usepackage{cuted}
\usepackage{hyperref}
\usepackage{amsmath}
\usepackage{cuted}
\usepackage{tabularx}
\usepackage{makecell}
\usepackage{amssymb} 
\usepackage{booktabs} 

\usepackage{tikz}
\usetikzlibrary{arrows.meta,positioning,fit,calc,matrix}
\usepackage{adjustbox}

\makeatletter
\renewcommand{\maketag@@@}[1]{\hbox{\m@th\normalsize\normalfont#1}}
\makeatother

{}
{}
{}
{}
{\bfseries}
{}
{ }
{}
\theoremstyle{theorembox}

\newtheorem{theorem}{Theorem}

\newtheorem{remark}{Remark}

\begin{document}

    \bstctlcite{BSTcontrol}
	
	\title{Multimodal Learning for MIMO Beam Prediction Based on Variational Inference}
	
	\author{Zijian Zheng, \IEEEmembership{Student Member, IEEE,}\
		Wenqiang Yi, \IEEEmembership{Member, IEEE,}
		Hyundong Shin, \IEEEmembership{Fellow, IEEE}\\
		and Arumugam Nallanathan, \IEEEmembership{Fellow, IEEE}
		\thanks{Zijian Zheng and Arumugam Nallanathan are with the School of Electronic Engineering and Computer Science, Queen Mary University of London, London, U.K. (emails: \{\href{mailto:z.zheng@qmul.ac.uk}{z.zheng}, \href{mailto:a.nallanathan@qmul.ac.uk}{a.nallanathan}\}@qmul.ac.uk).}
		\thanks{Wenqiang Yi is with the School of Computer Science and Electronic Engineering, University of Essex, Colchester CO4 3SQ, U.K. (email: \href{mailto:w.yi@essex.ac.uk}{w.yi@essex.ac.uk}).}
		\thanks{Hyundong Shin is with the Department of Electronics and Information Convergence Engineering, Kyung Hee University, Yongin-si, Gyeonggido 17104, Republic of Korea (e-mail: \href{hshin@khu.ac.kr}{hshin@khu.ac.kr}).}
	}
	
	\maketitle
	
	\begin{abstract}
        Accurate beam prediction is essential for mitigating signalling overhead and latency in integrated sensing and communication-enabled massive multi-input multi-output systems. With the aid of multimodal learning, the prediction accuracy can be enhanced by leveraging the complementary information from other existing sensors, but the practical deployment is often constrained by the high cost of acquiring semantically aligned multimodal datasets. This paper proposes a variational-inference-based multimodal framework that decouples the optimization problem into modular feature extraction and cross-modal semantic alignment. Specifically, we develop a two-stage training strategy where the model utilises abundant unimodal data for representation learning before performing refined alignment on limited multimodal samples. This design enhances data efficiency and ensures robust feature fusion under sensing uncertainties. Experimental results on the DeepSense6G dataset demonstrate that the proposed framework achieves competitive beam prediction accuracy and maintains high reliability, while only requiring $20\%$ of the multimodal training data compared to conventional end-to-end benchmarks.
	\end{abstract}
	
	
	\section{Introduction}

	The vision of sixth-generation (6G) mobile communication systems is to expand the boundary of communication infrastructures, evolving them from purely data-delivery pipelines into an integrated platform that jointly supports connectivity, sensing, and in-network intelligence\cite{NetOfEverything,isac-6g2}. In recent 6G surveys and outlooks, sensing and AI have been recognized as primary design elements for next-generation radio access and core network architectures. A practical motivation behind this vision is that current communication infrastructures, such as cellular base stations (BSs), have already been densely deployed; consequently, seamless wireless coverage has been achieved in many regions within the main service areas. Leveraging the inherent potential of wireless signals for sensing tasks, these ubiquitously deployed BSs offer a low-cost opportunity to build large-scale sensing networks without redundant infrastructure deployment\cite{ISAC}. In scenarios such as smart cities and intelligent transportation, reusing communication infrastructures as sensing devices has become a widely accepted premise.
	
	With the rapid development of multi-input multi-output (MIMO) techniques, radio frequency (RF) sensing under integrated sensing and communication (ISAC) frameworks has become a signature research direction, with a large body of work analyzing how the same spectrum, hardware, and waveforms can support both message delivery and environmental inference. However, sensing and communication pursue different objectives, and shared resources as well as mutual interference often couple their optimization goals. As a result, the sensing accuracy is not only constrained by non-ideal factors such as channel variations, multipath propagation, and dynamic blockages, but also inherits waveform- and protocol-related limitations from the communication system~\cite{ISAC}. Studies grounded in estimation theory and information theory suggest that, under fixed waveform designs and limited system degrees of freedom, improving the performance of one task in ISAC may lead to degradation in the other \cite{isac-limitation}. These observations motivate the adoption of multimodal learning as a fundamental approach for ISAC, particularly for sensing-aided beam prediction \cite{9857786}. By leveraging multimodal learning, beam prediction can exploit not only conventional estimated wireless channel state information, but also physical-world object direction cues obtained from heterogeneous sensors, thereby significantly improving prediction accuracy. Devices such as computer-vision sensors, LiDARs, and inertial measurement units (IMUs) can provide complementary information to the wireless modality as tested in \cite{zheng2024semi}; meanwhile, the wide-area coverage of wireless signals, their robustness under low-light conditions and visual occlusions, and their privacy-friendly characteristics can also benefit sensing with other modalities.
	
	In this research area, a straightforward multimodal starting point is \textit{early fusion}, where raw or lightly processed modality measurements are concatenated or otherwise merged at the input stage and fed into a single downstream network. A central benchmark enabling this line of work is the DeepSense6G dataset\cite{DeepSense6G} released by Alkhateeb and collaborators, which provides synchronized sensing and communication measurements in real-world scenarios and has become a common testbed for multimodal beam prediction and related tasks.
	
	While early fusion is conceptually simple, it tends to inherit practical weakness. Early fusion scheme forces raw or lightly processed inputs with incompatible sampling rates, noise statistics, and spatial semantics into a single representation before each modality has formed a stable, task-relevant description, which can amplify cross-modal interference rather than reduce uncertainty\cite{early-fusion-drawbacks}. \textit{Intermediate fusion}, also known as \textit{feature-level fusion}, mitigates this issue by first learning modality-specific feature extractors that respect each sensor’s inductive biases, and then performing feature-level interaction only after the representations have been normalized to comparable semantic granularity. For example, in \cite{Centerfusion}, the authors adopt a structured intermediate-variable fusion paradigm and highlight the key challenges in such frameworks, namely the design of appropriate alignment mechanisms and fusion operators. In \cite{Centerfusion}, the authors incorporate explicit geometric priors to constrain the intermediate variables produced by different modalities via frustum association and pillar expansion, and use concatenation as the fusion operator across modalities. Other representative designs include learnable alignment mechanisms implemented by semantic networks, like in \cite{Grif-Net, JSAC-gate-learnable}; gated aggregation operators that suppress unreliable modalities, for example in \cite{Grif-Net, Gate-Baseline}; and attention-based aggregation, e.g., \cite{TransformerFusion-Comm, Craft}.
	
	Given the challenges in designing suitable alignment mechanisms and selecting effective aggregation operators, an alternative for multimodal fusion is to further postpone fusion to the decision level, commonly referred to as \textit{late fusion} or \textit{decision-based fusion}. This strategy can be traced back to early signal-processing studies, such as the Chair-Varshney fusion rule \cite{c-v-rule}, and it remains widely adopted in modern multimodal settings. For instance, in \cite{late-fusion-early-work-comm}, the authors perform a second-stage detection using sub-6G wireless signals after the image modality reports a missed detection, thereby realizing late fusion in a dual-stage manner. In contrast to the sequential decision-revision pattern in \cite{late-fusion-early-work-comm}, the authors in \cite{vision-aided-6g-proactive-handoff} exemplify a parallel decision-combination scheme, where multiple base stations form their decisions independently and fusion occurs only when combining these decisions to trigger the final network action. Beyond the decision-thresholding schemes adopted in the two studies above, other works also employ soft decisions for multimodal aggregation, like \cite{omni-cnn,late-fusion-soft}.
	
	Across this line of designs, a recurring practical bottleneck is that most high-performance multimodal systems require end-to-end training on semantically aligned multimodal data, which needs costly collection and calibration in real-world wireless environments. More importantly, when the target scale is a wide-area communication network, such as an entire 6G system, continuously acquiring real-world multimodal data with stable semantic alignment becomes operationally and economically challenging, especially under privacy constraints and deployment limitations.

    \begin{figure}[t]
		\centering
		\includegraphics[width=0.8\linewidth]{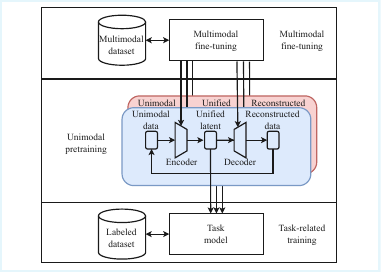}
		\caption{Workflow of the proposed multimodal framework: unimodal data are first used for unimodal self-supervised pretraining to reduce the dependence on multimodal data, followed by multimodal fine-tuning. Finally, a task model is trained separately to adapt to the target task.}
		\label{fig:system_chart}
	\end{figure}
	
	In this paper, we revisit the objective of multimodal learning from an optimization perspective. By adopting a variational-inference-based generative modeling framework and several specially designed regularization objectives, we develop a multimodal framework that supports modular training and development. In our framework, we employ a two-stage training paradigm: each unimodal model can be pretrained on abundant and inexpensive unimodal unlabeled data, and the pretrained models are then fine-tuned using a small amount of modality-aligned multimodal data to achieve modality alignment and cross-modal knowledge transfer. The major contributions of this work are summarized as follows:
    \begin{itemize}
        \item We propose a variational-inference-based modular multimodal framework for MIMO beam prediction in ISAC scenarios.
        \item We develop a regularization-based decoupling mechanism that separates unimodal feature learning from multimodal semantic alignment, thereby enabling modular development, training, and deployment of modality-specific sub-models.
        \item We design a unified latent space with a shared--private dual-latent structure, which facilitates effective extraction of modality-private features while enabling robust alignment of modality-invariant representations.
        \item We introduce a temporal modeling mechanism that equips the framework with the ability to capture not only single-frame features but also temporal dynamics in sequential signals.
        \item Experiments on the DeepSense6G\cite{DeepSense6G} dataset validate the effectiveness of the proposed framework. With only $20\%$ of modality-aligned multimodal data, our framework achieves performance comparable to the benchmarks; when all multimodal data are available, it outperforms the benchmarks in two scenarios.
    \end{itemize}

    The rest of this paper is organized as follows. In Section \ref{sec:sys_model}, we introduce the system model and formulate the optimization objective. In Section \ref{sec:alg_design}, we further analyze the optimization problem and progressively develop the corresponding algorithmic framework. In Section \ref{sec:model_design}, we present the design details of the subsystem models involved in the framework. The experimental setup and performance evaluation are provided in Section \ref{sec:experiments}. Finally, Section \ref{sec:conclusion} concludes the paper.

	\section{System Model and Problem Formulation}
	\label{sec:sys_model}
	This section first presents the considered MIMO system for ISAC, which mainly focuses on angular estimation based on the received signal strength. To enhance the sensing accuracy under ISAC-enabled networks, the studied multimodal sensor fusion problem is formally formulated by considering the modality correlations.
	\subsection{System Model}
	\label{subsec:rf_model}
	Thanks to the ubiquitous coverage of cellular BSs, RF signals can be the primary sensing modality to perceive the blind spots of user equipment (UE). It provides complementary information to traditional on-board sensors. In our model, both BSs and UE are assumed to be equipped with MIMO capabilities. The BS is configured with $N_{tx}$ transmit antennas, while the UE has $N_{rx}$ receive antennas. The channel matrix between the UE and the BS can be characterized via a geometrically sparse multipath model \cite{channel_model} with $L$ multipaths,
	\begin{equation}
		\mathbf H \in \mathbb C^{N_{rx} \times N_{tx}} = \sum_{l=1}^L \alpha_l \mathbf a_{rx}(\theta_l) \mathbf a_{tx}(\phi_l)^\mathrm{H},
	\end{equation}
	where $\alpha_l \in \mathbb C$ is the complex channel gain of the $l$-th path, $\theta_l$ is the angle of arrival (AoA), $\phi_l$ is the angle of departure (AoD), $\mathbf a_{rx}(\cdot)$ and $\mathbf a_{tx}(\cdot)$ are the array steering vectors. We assume a narrowband channel, meaning that the channel coefficients remain constant across all subcarriers within each symbol. Therefore, the channel coefficient corresponding to the $n$-th symbol is denoted as $\mathbf H[n] \in \mathbb C^{N_{rx} \times N_{tx}}$.
	
	Let the training symbol be denoted as $s[n]$, and its energy is expressed as:
	\begin{equation}
		P_s = \mathbb E \left( \left\| s[n] \right\|^2 \right),
	\end{equation}
	and the receive vector is given by:
	\begin{equation}
		\mathbf y[n] \in \mathbb C^{N_{rx}} = \mathbf H[n] \mathbf f[n] s[n] + \mathbf v[n],
	\end{equation}
	where $\mathbf f[n] \in \mathbb C^{N_{tx}}$ is the transmit precoder and $\| f[n] \|_2 = 1$, $\mathbf v[n] \sim \mathcal{CN}(\mathbf 0, \sigma_v^2\mathbf I)$ is the additional white Gaussian noise (AWGN).
	
	Let the number of codewords at the receiver be $B$, and denote $\mathbf g_b \in \mathbb C^{N_{rx}}$ as the $b$-th receive combiner which satisfies $\left\| \mathbf g_b \right\|_2 = 1$. Then, the signal received by the $b$-th beam at the $n$-th symbol can be expressed as:
	\begin{equation}
		z_b[n] = \mathbf g_b^\mathrm{H} \mathbf y[n] = \mathbf g_b^\mathrm{H}\mathbf H[n] \mathbf f[n] s[n] + \mathbf g_b^\mathrm{H}\mathbf v[n].
	\end{equation}
	
	With the receive vectors from each beam, we can estimate the received power vector as:
	\begin{equation}
		\mathbf p[n] = \begin{bmatrix}
			\hat{P_1} & \cdots & \hat{P_b} & \cdots & \hat{P_B}
		\end{bmatrix}^\mathrm T \in \mathbb R_+^B,
	\end{equation}
	where $\hat{P_b}$ is the estimated power from the $b$-th beam, denoted as:
	\begin{equation}
		\hat{P_b} = \frac{1}{\mathcal N} \sum_{n=1}^\mathcal N \left| z_b[n] \right|^2,
	\end{equation}
	where $\mathcal N$ is the size of the window.

	\subsection{Problem Formulation}
	\label{subsec:joint_opt_description}
	
	Note that the data collected from the aforementioned MIMO system is only one modality based on RF signals, which highlights some characteristics of the sensed environment, but lacks high-resolution and scene-level information. Therefore, we leverage the complementary information from existing on-board sensors to further enhance the sensing accuracy. We denote the raw data inputs of the $M$ modalities as $(\mathbf x_1, \cdots, \mathbf x_M)$, where $\mathbf x_1$ represents the data from RF sensing and $(\mathbf x_2, \cdots, \mathbf x_m,\cdots, \mathbf x_M)$ are the data from the rest on-board sensors. Different sensors have different data structures. For example, when $\mathbf x_m$ represents the GNSS modality, $\mathbf x_m = [(x_1, y_1, z_1)^\mathrm T, (x_2, y_2, z_2)^\mathrm T, \cdots]$ corresponds to a time series of world coordinates obtained from multiple GNSS receivers. When $\mathbf x_m$ represents the RF modality, $\mathbf x_m = [\mathbf B[1], \mathbf B[2], \cdots]$, where $\mathbf B[n] \in \mathbb C^{N_{rx} \times S \times C}$ denotes the RF data cube of the $n$-th frame. Here, $N_{rx}$ is the number of receive antennas, $S$ is the number of samples within each chirp, and $C$ is the number of chirps per frame.
	
	Next, we discuss the considered multimodal sensor fusion problem with the aid of deep learning schemes. Consider a supervised multimodal learning problem with $M$ heterogeneous modalities. For the $i$-th sample, the multimodal observation is denoted by
	\begin{equation}
		\mathbf X^{(i)} = \big(\mathbf x_1^{(i)}, \cdots, \mathbf x_M^{(i)}\big),
	\end{equation}
	and the corresponding label is $\mathbf y^{(i)}$. A discriminative model $f(.,.)$ parameterized by $\mathbf w$ produces a prediction
	\begin{equation}
		\hat{\mathbf y}^{(i)} = f(\mathbf w, \mathbf X^{(i)}).
	\end{equation}
	Given a loss function $L(\cdot,\cdot)$, parameter estimation follows the empirical risk minimization principle:
	\begin{equation}
		\mathbf w^{*} = \arg\min_{\mathbf w} \frac{1}{D} \sum_{i=1}^{D} L\big(\mathbf y^{(i)}, \hat{\mathbf y}^{(i)}\big),
	\end{equation}
	where $D$ is the size of the dataset. In this formulation, the input space is the Cartesian product of the modality-specific spaces, and all modalities jointly influence the update of the shared parameter vector $\mathbf w$. The gradient of the loss with respect to a modality-specific block $\mathbf w_m$ satisfies
	\begin{equation}
		\nabla_{\mathbf w_m} L\big(\mathbf y^{(i)}, \hat{\mathbf y}^{(i)}\big) =\frac{\partial L}{\partial f} \frac{\partial f(\mathbf w, \mathbf X^{(i)})}{\partial \mathbf w_m},
	\end{equation}
	and the Jacobian $\partial f(\mathbf w, \mathbf X^{(i)}) / \partial \mathbf w_m$ generally depends on all components of $\mathbf X^{(i)}$. Thus, different modalities remain coupled through a unified parameter set and a shared optimization objective, which limits modularity and impedes training when modalities are missing or only partially available.
	
	To obtain a more structured view of multimodal interactions, we introduce a latent variable $\mathbf z$ representing shared semantic factors underlying all modalities. With this latent representation, the statistical relationships between modalities and labels can be organized through a structured joint model. Specifically, we consider the latent-variable generative model
	\begin{equation}
		p_{\theta}(\mathbf X, \mathbf y, \mathbf z) = p(\mathbf z) p_{\theta}(\mathbf y | \mathbf z) \prod_{m=1}^{M} p_{\theta}(\mathbf x_m | \mathbf z),
		\label{eq:joint_split}
	\end{equation}
	where $p(\mathbf z)$ is a prior distribution over the latent space, and $p_{\theta}(\mathbf x_m | \mathbf z)$ and $p_{\theta}(\mathbf y | \mathbf z)$ describe the conditional distributions of the modalities and the label, respectively. Here $\theta$ represents the parameter for the parameterized probability. The factorization expresses conditional independence of all observed variables given $\mathbf z$, making $\mathbf z$ an explicit mediating representation that organizes multimodal dependencies.
	
	Under this probabilistic model, parameter estimation follows the maximum likelihood principle:
	\begin{equation}
		\theta^{*} = \arg\max_{\theta} \log p_{\theta}(\mathbf X, \mathbf y) = \arg\max_{\theta} \log \int p_{\theta}(\mathbf X, \mathbf y, \mathbf z) \mathrm d\mathbf z.
	\end{equation}
	Direct computation of the integral over $\mathbf z$ is generally intractable, which necessitates the introduction of a tractable surrogate objective function. To this end, we introduce a parameterized variational distribution with parameters $\phi$, denoted by $q_{\phi}(\mathbf z | \mathbf X, \mathbf y)$. Using the identity
	\begin{equation}
		\log p_{\theta}(\mathbf X, \mathbf y) = \log \int q_{\phi}(\mathbf z | \mathbf X, \mathbf y) \frac{p_{\theta}(\mathbf X, \mathbf y, \mathbf z)}{q_{\phi}(\mathbf z | \mathbf X, \mathbf y)} \mathrm d\mathbf z,
	\end{equation}
	and applying Jensen’s inequality yields a variational lower bound on the log-likelihood:
	\begin{equation}
		\log p_{\theta}(\mathbf X, \mathbf y) \ge \int q_{\phi}(\mathbf z | \mathbf X, \mathbf y)
		\log \frac{p_{\theta}(\mathbf X, \mathbf y, \mathbf z)}{q_{\phi}(\mathbf z | \mathbf X, \mathbf y)} \mathrm d\mathbf z.
	\end{equation}
	
	Motivated by this inequality, we adopt the following lower bound as the optimization objective function:
	\begin{equation}
		\mathcal L(\theta, \phi) =  \mathbb E_{q_{\phi}(\mathbf z | \mathbf X, \mathbf y)} \left[ \log p_{\theta}(\mathbf X, \mathbf y, \mathbf z) - \log q_{\phi}(\mathbf z | \mathbf X, \mathbf y) \right].
	\end{equation}
	Maximizing this bound with respect to both $\theta$ and $\phi$ provides a tractable approximation to maximum likelihood estimation:
	\begin{equation}
		(\theta^{*}, \phi^{*}) = \arg\max_{\theta, \phi} , \mathcal L(\theta, \phi).
	\end{equation}
	
	Lastly, substituting the factorized form of $p_{\theta}(\mathbf X, \mathbf y, \mathbf z)$ in equation \eqref{eq:joint_split} leads to
	\begin{equation}
		\begin{aligned}
			\mathcal L(\theta, \phi) &= \underbrace{\mathbb E_{q_{\phi}} \left[ \log p_{\theta}(\mathbf y | \mathbf z) \right]}_{(a)} + \sum_{m=1}^{M} \underbrace{\mathbb E_{q_{\phi}} \left[ \log p_{\theta}(\mathbf x_m | \mathbf z) \right]}_{(b)}\\
			&\quad - \underbrace{D_{\mathrm{KL}} \left[ q_{\phi}(\mathbf z | \mathbf X, \mathbf y) \| p(\mathbf z) \right]}_{(c)}.
		\end{aligned}
		\label{eq:lb}
	\end{equation}
	The derived objective function in \eqref{eq:lb} consists of expected likelihood contributions from the label and each modality, denoted as the terms (a) and (b) in the equation, together with a regularization term that encourages the approximate posterior to remain close to the prior denoted as term (c). All expectations are computed with respect to the same posterior distribution $q_{\phi}(\mathbf z | \mathbf X, \mathbf y)$, which aggregates information across all modalities and the supervisory variables. Consequently, parameter updates for both $\theta$ and $\phi$ are driven by a shared latent representation, providing a principled account of cross-modal interactions within the latent-variable framework.
	
	\section{Algorithm Design}
	\label{sec:alg_design}
	In this section, we present our multimodal sensor fusion algorithm. We first present a generalized unified framework based on variational inference. Building on this framework, we progressively introduce designs that facilitate inter-modality decoupling and alignment, analyze its extension to temporal modeling, and finally derive the resulting optimization objective after incorporating these designs.
	\subsection{Generative Model and Variational Inference}
	\label{subsec:gen_vi}
	
	When multiple modalities are introduced, the optimization problem defined in \ref{sec:sys_model} becomes highly coupled. Both the objective function and the optimization variables across different modalities are mutually related. This coupling results in the curse of dimensionality, which hinders modular learning and efficient training. To obtain a tractable objective while preserving an explicit probabilistic structure, we introduce a latent-variable generative model and perform inference via variational approximation. This converts the coupled optimization into a marginal likelihood maximization problem, where the intractable marginalization over latent variables is handled by the evidence lower bound (ELBO).
	
	Let $\mathcal{M}=\{1,\cdots,M\}$ denote the set of modalities and $\mathcal{T}=\{1,\cdots,T\}$ denote the time index. Our multimodal observations $X_{1:T}$ and ground-truth values $Y_{1:T}$ construct two sequences, which are
	\begin{equation}
		X_{1:T} \triangleq \{x_t^m\}_{t\in\mathcal{T}, m\in\mathcal{M}},\;
		Y_{1:T} \triangleq \{y_t\}_{t\in\mathcal{T}},
	\end{equation}
	where $x_t^m$ denotes the sampled data at time $t$ and the modality $m$. The $y_t$ denotes the task output at time $t$, either a classification tag or prediction value. To capture cross-modal shared information, modality-specific complementary knowledge, and temporal context, we introduce a shared latent variable $z_t^{\mathrm{s}}$, a modality-private latent variable $z_t^{\mathrm{p},m}$, and an auxiliary state variable $h_t$, respectively. As the input time index evolves, these variables naturally form time series, which we denote as:
	\begin{equation}
		\begin{aligned}
			Z_{1:T}^{\mathrm{s}} &\triangleq \{z_t^{\mathrm{s}}\}_{t\in\mathcal{T}},\\
			Z_{1:T}^{\mathrm{p}} &\triangleq \{Z_t^{\mathrm{p}}\}_{t\in\mathcal{T}},\\
			H_{0:T} &\triangleq \{h_0\} \cap \{h_{t}\}_{t\in\mathcal{T}},
		\end{aligned}
	\end{equation}
	where $Z_t^{\mathrm{p}} \triangleq \{z_t^{\mathrm{p},m}\}_{m\in\mathcal{M}}$ is a set rather than a variable, because the modality-private latent variables differ across modalities. We explicitly list $h_0$ because, unlike the other variables, $h_0$ denotes the initial system state before any observations are available, rather than a representation induced by observations at $t>0$. After defining the necessary notations, we introduce the considered generative model and variational inference separately. 
	
	\subsubsection{Unified Generative Model}
		Let $h_{t-1}$ be a sufficient statistic at time $t$, i.e., the latent $z_t^\mathrm s$, $z_t^{\mathrm p,m}$	and state variable $h_t$ satisfy a first-order Markov assumption, and the latent and state variables constitute another sufficient statistic of the observations $x_t^m$. Then, by the chain rule, we partition the sequential variables into two groups, namely the intermediate variables $z^{\mathrm s}_t, Z^{\mathrm p, m}_t, h_t$ and the observations $X_t$, and expand the joint distribution along $t$, yielding:
	\begin{equation}
		\begin{aligned}
			&p_\theta\left(y, X_{1:T}, Z_{1:T}^{\mathrm{s}}, Z_{1:T}^{\mathrm{p}}, H_{0:T}\right) \\
			&= p_\theta(h_0)\; p_\theta\left(z_1^{\mathrm{s}}, Z_1^{\mathrm{p}}, h_1 | h_0\right)\; p_\theta\left(X_1 | h_0, z_1^{\mathrm{s}}, Z_1^{\mathrm{p}}, h_1\right) \\
			&\quad\times \prod_{t=2}^{T} p_\theta\left(z_t^{\mathrm{s}}, Z_t^{\mathrm{p}}, h_t | h_0,\{(z_\tau^{\mathrm{s}}, Z_\tau^{\mathrm{p}}, h_\tau)\}_{\tau=1}^{t-1}, X_{1:t-1}\right) \\
			&\quad\times \prod_{t=2}^{T} p_\theta\left(X_t | h_0,\{(z_\tau^{\mathrm{s}}, Z_\tau^{\mathrm{p}}, h_\tau)\}_{\tau=1}^{t}, X_{1:t-1}\right) \\
			&\quad\times p_\theta\left(y | h_0,\{(z_\tau^{\mathrm{s}}, Z_\tau^{\mathrm{p}}, h_\tau)\}_{\tau=1}^{T}, X_{1:T}\right).
		\end{aligned}
		\label{eq:chain_rule}
	\end{equation}
	
	Under the first-order Markov assumption, conditioned on $h_{t-1}$, the current-step latent variables $z_t^\mathrm s, Z_t^\mathrm p$ and observations $X_t$ are independent of earlier time indices, and each modality is generated conditionally independently given $(z_t^\mathrm s, z_t^{\mathrm p,m}, h_{t-1})$. Moreover, $y$ depends on the multimodal sequence only through $(Z_{1:T}^\mathrm s, H_{0:T})$. Thus, \eqref{eq:chain_rule} reduces to:
	\begin{equation}
		\begin{aligned}
			&p_\theta\left(y, X_{1:T}, Z_{1:T}^{\mathrm{s}}, Z_{1:T}^{\mathrm{p}}, H_{0:T}\right)\\
			&\quad= p(h_0)\prod_{t=1}^{T} p_{\theta_m,\theta_y,\vartheta}\left(z_t^{\mathrm{s}}, Z_t^{\mathrm{p}}, h_t | h_{t-1}\right) \\
			&\quad\quad\times \prod_{t=1}^{T}\prod_{m\in\mathcal{M}}
			p_{\theta_m}\left(x_t^m | z_t^{\mathrm{s}}, z_t^{\mathrm{p},m}, h_{t-1}\right)\;
			p_{\theta_y}\left(y | Z_{1:T}^{\mathrm{s}}, H_{0:T}\right).
		\end{aligned}
		\label{eq:unified_model}
	\end{equation}	
	where $\theta=\{\theta_y,\{\theta_m\}_{m\in\mathcal{M}},\vartheta\}$ collects all learnable parameters. In particular, here $\vartheta$ parameterizes the state transition $p_\vartheta(h_t | h_{t-1})$.
	
	The learning objective can be cast as maximization of the marginal log-likelihood:
	\begin{equation}
		\begin{aligned}
			&\log p_\theta(y, X_{1:T})\\
			&\; =
			\log \int
			p_\theta\left(y, X_{1:T}, Z_{1:T}^{\mathrm{s}}, Z_{1:T}^{\mathrm{p}}, H_{0:T}\right)
			\mathrm dZ_{1:T}^{\mathrm{s}} \mathrm dZ_{1:T}^{\mathrm{p}} \mathrm dH_{0:T},
		\end{aligned}
	\end{equation}
	which is generally intractable due to the high-dimensional integral over
	$\left(Z_{1:T}^{\mathrm{s}}, Z_{1:T}^{\mathrm{p}}, H_{0:T}\right)$.
	
	\subsubsection{Variational Inference and ELBO}
	To obtain a tractable objective, we introduce a variational distribution generated by an inference network parameterized by $\phi$:
	\begin{equation}
		q_\phi\left(Z_{1:T}^{\mathrm{s}}, Z_{1:T}^{\mathrm{p}}, H_{0:T} | X_{1:T}, y\right).
	\end{equation}
	We further define
	\begin{equation}
		U \triangleq \left(Z_{1:T}^{\mathrm{s}}, Z_{1:T}^{\mathrm{p}}, H_{0:T}\right),
		\label{eq:define_U}
	\end{equation}
	as a shorthand that collects all latent variables and state values, so that
	\begin{equation}
		p_\theta(y, X_{1:T}) = \int p_\theta(y, X_{1:T}, U)\mathrm dU.
	\end{equation}
	By multiplying and dividing the integrand by $q_{\phi}(U | X_{1:T}, y)$ and rewriting the marginalization as an expectation, we have:
	\begin{equation}
		\begin{aligned}
			\log p_\theta(y, X_{1:T}) &= \log \int p_\theta(y, X_{1:T}, U) \mathrm dU\\
			&= \log \int q_\phi(U|X_{1:T},y) \frac{p_\theta(y, X_{1:T}, U)}{q_\phi(U|X_{1:T},y)}\mathrm dU\\
			&=\log \mathbb E_{q_\phi(U|X_{1:T},y)} \left[ \frac{p_\theta(y, X_{1:T}, U)}{q_\phi(U|X_{1:T},y)} \right].
		\end{aligned}
	\end{equation}
	Since $\log(\cdot)$ is concave, we apply the Jensen's inequality and end up with:
	\begin{equation}
		\begin{aligned}
			\log p_\theta(y, X_{1:T}) &\ge \mathbb E_{q_\phi(U|X_{1:T},y)} \log \left[ \frac{p_\theta(y, X_{1:T}, U)}{q_\phi(U|X_{1:T},y)} \right]\\
			&= \mathbb E_{q_\phi(U|X_{1:T},y)} \log \left[ p_\theta(y, X_{1:T}, U) \right]\\
			&\quad - \mathbb E_{q_\phi(U|X_{1:T},y)} \log \left[ q_\phi(U|X_{1:T},y) \right]\\
			&\triangleq \mathcal{L}(\theta,\phi),
			\label{eq:ELBO}
		\end{aligned}
	\end{equation}
	where $\mathcal{L}(\theta,\phi)$ is commonly named the evidence lower bound (ELBO) in variational Bayesian methods.
	
	Using the definition in \eqref{eq:define_U}, we have:
	\begin{equation}
		\begin{aligned}
			p_\theta(U)
			&= p_\theta \left(Z_{1:T}^{\mathrm s}, Z_{1:T}^{\mathrm p}, H_{0:T}\right)\\
			&= p_\theta(h_0) p_\theta \left(Z_{1:T}^{\mathrm s}, Z_{1:T}^{\mathrm p}, H_{1:T}| h_0\right)\\
			&\overset{(a)}{=} p(h_0)\prod_{t=1}^{T} p_\theta \left( z_t^{\mathrm s}, Z_t^{\mathrm p}, h_t | Z_{1:t-1}^{\mathrm s}, Z_{1:t-1}^{\mathrm p}, H_{0:t-1} \right)\\
			&\overset{(b)}{=} p(h_0)\prod_{t=1}^{T} p_\theta \left( z_t^{\mathrm s}, Z_t^{\mathrm p}, h_t | h_{t-1} \right),
		\end{aligned}
		\label{eq:prob_u}
	\end{equation}
	where $(a)$ follows from the chain rule, and $(b)$ follows from the first-order Markov assumption on the state variable $h_t$.
	
	Using the factorization in \eqref{eq:unified_model} and substituting \eqref{eq:prob_u}, the logarithm joint distribution admits
	\begin{equation}
		\begin{aligned}
			&\log p_\theta(y, X_{1:T}, U)\\
			&\quad = \log p_{\theta_y}\left(y \big| Z_{1:T}^{\mathrm{s}}, H_{0:T}\right) + \log p_\theta(U)\\
			&\quad\quad + \sum_{t=1}^{T}\sum_{m\in\mathcal{M}}
			\log p_{\theta_m}\left(x_t^m \big| z_t^{\mathrm{s}}, z_t^{\mathrm{p},m}, h_{t-1}\right).
		\end{aligned}
		\label{eq:simplified_unified_model}
	\end{equation}
	Substituting \eqref{eq:simplified_unified_model} into \eqref{eq:ELBO} and using linearity of expectation, we obtain
	\begin{equation}
		\begin{aligned}
			&\mathcal{L}(\theta,\phi)\\
			&\quad= \mathbb E_{q_\phi(U|X_{1:T},y)} \left[ \log p_{\theta_y}\left(y \big| Z_{1:T}^{\mathrm{s}}, H_{0:T}\right) \right]\\
			&\quad + \mathbb E_{q_\phi(U|X_{1:T},y)} \left[ \log p_\theta(U) \right]\\
			&\quad - \mathbb E_{q_\phi(U|X_{1:T},y)} \log \left[ q_\phi(U|X_{1:T},y) \right]\\
			&\quad + \sum_{t=1}^{T}\sum_{m\in\mathcal{M}} \mathbb E_{q_\phi(U|X_{1:T},y)} \left[ \log p_{\theta_m}\left(x_t^m \big| z_t^{\mathrm{s}}, z_t^{\mathrm{p},m}, h_{t-1}\right) \right]
		\end{aligned}
	\end{equation}
	Finally, by the definition of KL divergence, we equivalently have
	\begin{equation}
		\begin{aligned}
			&\mathbb E_{q_\phi} \left[ \log p_\theta(U) \right] - \mathbb E_{q_\phi} \left[ \log q_\phi(U|X_{1:T},y) \right]\\
			&\quad = - D_{\mathrm{KL}} \left( q_\phi\left(U | X_{1:T}, y\right) \big\| p_\theta(U) \right).
		\end{aligned}
	\end{equation}
	Therefore, $\mathcal{L}(\theta,\phi)$ can be written as
	\begin{equation}
		\begin{aligned}
			&\mathcal{L}(\theta,\phi)\\
			&\quad= \mathbb{E}_{q_\phi(U| X_{1:T},y)} \left[ \log p_{\theta_y}\left(y | Z_{1:T}^{\mathrm{s}}, H_{0:T}\right) \right] \\
			&\quad+ \sum_{t=1}^{T}\sum_{m\in\mathcal{M}} \mathbb{E}_{q_\phi(U| X_{1:T},y)} \left[ \log p_{\theta_m}\left(x_t^m | z_t^{\mathrm{s}}, z_t^{\mathrm{p},m}, h_{t-1}\right) \right] \\
			&\quad- D_{\mathrm{KL}} \left( q_\phi\left(U | X_{1:T}, y\right) \big\| p_\theta(U) \right),
		\end{aligned}
		\label{eq:unified_elbo}
	\end{equation}
	Maximizing the ELBO with respect to $(\theta,\phi)$ provides a tractable approach for the intractable marginal likelihood maximization. The expected log-likelihood terms decompose over modalities and time indices, while the KL divergence term regularizes the variational posterior toward the structured prior.

    \begin{remark}
	The unified formulation in \eqref{eq:unified_elbo} covers multiple scenarios. When $T=1$, the framework degenerates to a static multimodal formulation that does not rely on temporal context. When $M=1$, it reduces to a VRNN\cite{VRNN}, a unimodal dynamic variational inference model. When $T=1$ and $M=1$, it further degenerates to the conventional unimodal VAE\cite{VAE}.
    \end{remark}
    
    The maximization of the objective function in \eqref{eq:unified_elbo} is still challenging. We will reformulate it into a series of sub-questions: 1) Static inter-modality decoupling; 2) Private and shared latent fusing; and 3) Temporal feature coupling. They are discussed in the following subsections.

	\subsection{Multimodal Framework With Decoupled Alignment}
	\label{subsec:mm_shared}
	
	To explore the design principles for achieving inter-modality decoupling, we start with a static multimodal setting with a single shared latent representation without modality-private latents or temporal state. Specifically, we set $T=1$ and $\left\{ z_{1}^{\mathrm{p},m} \right\}_m$ are marginal independent with the observation $X$, thereby removing them from all probabilistic models. Denoting $x^m \triangleq x_1^m$ and $X \triangleq \{x^m\}_{m\in\mathcal{M}}$, the unified joint model in \eqref{eq:unified_model} reduces to
	\begin{equation}
		p_\theta(y,X,z^{\mathrm{s}}) = p(z^{\mathrm{s}}) p_{\theta_y}(y| z^{\mathrm{s}}) \prod_{m\in\mathcal{M}} p_{\theta_m}(x^m| z^{\mathrm{s}}).
		\label{eq:simpliest_prob_model}
	\end{equation}
	
	The specialization shown in \eqref{eq:simpliest_prob_model} makes $z^{\mathrm{s}}$ the only interface through which different modalities can interact. To make this design feasible in practice, we require $z^{\mathrm{s}}$ to satisfy certain statistical properties. In particular:	
	\subsubsection{Generative Model Design}
		For the generative model, $z^{\mathrm{s}}$ should be rich enough to account for all inter-modality statistical dependence. This requirement has two practical implications. First, once $z^{\mathrm{s}}$ is given, each modality should have its own modality-specific generator that does not depend on the other modalities, i.e.,
		\begin{equation}
			\begin{aligned}
				p_\theta(X|z^{\mathrm{s}}) = \prod_{m\in\mathcal{M}} p_{\theta_m}(x^m|z^{\mathrm{s}}).
			\end{aligned}
			\label{eq:dec_factor_static}
		\end{equation}
		Equation \eqref{eq:dec_factor_static} imposes a \emph{factorized decoder}: conditioned on the shared variable $z^{\mathrm{s}}$, the joint conditional distribution over all modalities decomposes into a product of modality-wise terms. Equivalently, it enforces a conditional-independence structure:
		\begin{equation}
			x^{m} \perp\!\!\!\perp x^{m'} \;\big|\; z^{\mathrm{s}}, \quad \forall m\neq m'.
			\label{eq:condit_inden}
		\end{equation}
		As a result, any cross-modal statistical dependence must be expressed through $z^{\mathrm{s}}$, while any remaining uncertainty is modeled within each modality-specific factor $p_{\theta_m}(x^m | z^{\mathrm{s}})$. Practically, this factorization rules out direct cross-modal connections in the decoder and allows each term $p_{\theta_m}(x^m | z^{\mathrm{s}})$ to be implemented as an independent modality expert.
		
		Second, \eqref{eq:dec_factor_static} implies an approach to achieve Bayesian evidence aggregation. By Bayes' rule, the exact posterior of $z^{\mathrm{s}}$ given all modalities is
		\begin{equation}
			\begin{aligned}
				p_\theta(z^{\mathrm{s}}|X)
				&=
				\frac{
					p(z^{\mathrm{s}})\,p_\theta(X|z^{\mathrm{s}})
				}{
					p_\theta(X)
				}\\
				&=
				\frac{
					p(z^{\mathrm{s}})\,p_\theta(X|z^{\mathrm{s}})
				}{
					\int p(z^{\mathrm{s}})\,p_\theta(X|z^{\mathrm{s}})\mathrm dz^{\mathrm{s}}
				}.
			\end{aligned}
			\label{eq:bayes_rule_full}
		\end{equation}
		Substituting the conditional-independence design \eqref{eq:dec_factor_static} into \eqref{eq:bayes_rule_full} yields
		\begin{equation}
			\begin{aligned}
				p_\theta(z^{\mathrm{s}}|X)
				&=
				\frac{
					p(z^{\mathrm{s}})\,\prod_{m\in\mathcal{M}} p_{\theta_m}(x^m|z^{\mathrm{s}})
				}{
					\int p(z^{\mathrm{s}})\,\prod_{m\in\mathcal{M}} p_{\theta_m}(x^m|z^{\mathrm{s}})\mathrm dz^{\mathrm{s}}
				}\\
				&\propto
				p(z^{\mathrm{s}})
				\prod_{m\in\mathcal{M}} p_{\theta_m}(x^m|z^{\mathrm{s}}),
			\end{aligned}
			\label{eq:bayes_posterior_product}
		\end{equation}
		which shows that each modality contributes a multiplicative likelihood term to the shared latent posterior under a common prior.
		
		Although in \eqref{eq:unified_model} we specialize our model to a factorized likelihood family from the outset, the conditional-independence requirement in \eqref{eq:condit_inden} is ultimately a property of the learned conditional distributions and is not guaranteed to hold automatically in practice. In particular, if the shared latent variable $z^{\mathrm{s}}$ is not sufficiently expressive to absorb all cross-modal dependence, the learned likelihoods may still exhibit residual conditional dependence.
		
		In principle, one may explicitly penalize conditional dependence by minimizing a differentiable conditional independence measure, e.g., the conditional mutual information $\mathcal{I}(x^{m};x^{m'} | z^{\mathrm{s}})$ estimated by neural mutual-information estimators like MINE\cite{MINE,c-MINE}, or kernel-based conditional independence criteria such as HSIC\cite{hsic, c-hsic}. However, obtaining stable gradients from such estimators is often computationally expensive and sensitive to architecture and hyper-parameter choices. We therefore adopt an engineering approach that encourages approximate conditional independence via decorrelation.
		
		Let $\hat{x}^{m}_{\theta}(z^{\mathrm{s}}) \triangleq \mathbb{E}_{p_{\theta_m}(x^{m}| z^{\mathrm{s}})}(x^{m})$ denote the mean of the likelihood $p_{\theta_m}(x^{m}| z^{\mathrm{s}})$. We define the residual feature as
		\begin{equation}
			r_{m} \triangleq x^{m} - \hat{x}^{m}_{\theta}(z^{\mathrm{s}}), \quad z^{\mathrm{s}} \sim q_{\phi}(z^{\mathrm{s}}| X).
			\label{eq:dec_residual_feat}
		\end{equation}

		We use the residual $r_m$, rather than the reconstruction $\hat{x}^{m}_{\theta}(z^{\mathrm{s}})$, as the operand for the decorrelation because $\hat{x}^{m}_{\theta}(z^{\mathrm{s}})$ is generated conditionally on latent $z^{\mathrm{s}}$ and therefore does not reflect the stochasticity induced by $z^{\mathrm{s}}$. As a result, decorrelating via $\hat{x}^{m}_{\theta}(z^{\mathrm{s}})$ effectively attempts to decorrelate the prior of the raw data, which may undermine the representational ability of the model and is not our objective. In contrast, using the residual provides an appropriate measure of correlation in the conditional likelihood.
		
		If the decoder-side conditional independence holds, then the remaining uncertainty beyond $z^{\mathrm{s}}$ should be modality-local, and the cross-modal dependence in $\{r_{m}\}_m$ should be suppressed. Accordingly, we impose the following regularization:
		\begin{equation}
			\mathcal{R}_{\mathrm{dec}}(\theta) \triangleq \sum_{\substack{m,m'\in\mathcal{M}\\ m<m'}} \mathbb{E}_{(X,y)} \left[ \left\| \widehat{\mathrm{Cov}} \left( r_{m}, r_{m'} \right) \right\|_F^2 \right],
			\label{eq:reg_dec_static}
		\end{equation}
		where $\widehat{\mathrm{Cov}}(\cdot,\cdot)$ denotes the empirical cross-covariance estimated within a mini-batch.
		
		Minimizing $\mathcal{R}_{\mathrm{dec}}$ suppresses correlations among the modality-specific likelihoods as an engineering approximation to enhance independence, thereby driving our model toward our designed theoretical framework. To distinguish it from another regularization term introduced later, we refer to this term as \emph{decoder regularization}. This naming is motivated by the fact that it acts on the sub-network whose learning target is the likelihood $p_{\theta_m}(x^m|z^{\mathrm{s}})$, which is commonly referred to as the \emph{decoder} in VAE-style architectures.
		
		\subsubsection{Inference Model Design}
        The inference model should map different modalities into a common latent space such that the resulting latent representations are mutually compatible across modalities.
		To this end, we introduce modality-specific variational posteriors $q_{\phi_m}(z^{\mathrm{s}}|x^m)$ and construct a fused posterior by a product-of-experts (PoE) aggregation,
		\begin{equation}
			\begin{aligned}
				q_\phi(z^{\mathrm{s}}|X)
				&=
				\frac{1}{\mathcal{Z}(X)}
				p(z^{\mathrm{s}})
				\prod_{m\in\mathcal{M}}
				q_{\phi_m}(z^{\mathrm{s}}|x^m)^{\alpha_m},
			\end{aligned}
			\label{eq:mm_poe}
		\end{equation}
		where $\alpha_m\ge 0$ are learnable reliability weights and $\mathcal{Z}(X)$ is the normalizer. This structure is consistent with the product form in \eqref{eq:bayes_posterior_product} and naturally supports missing modalities by omitting absent experts.
		
		However, without additional constraints, unimodal encoders may learn incompatible latent semantics, which makes cross-modality aggregating unstable. Therefore, we enforce a posterior-consistency regularization that aligns each unimodal posterior with the fused posterior:
		\begin{equation}
			\begin{aligned}
				&\mathcal{R}_{\mathrm{enc}}(\phi)\\
				&\quad \triangleq \sum_{m\in\mathcal{M}} \mathbb{E}_{(X,y)} \left[ D_{\mathrm{KL}} \left( q_{\phi_m}(z^{\mathrm{s}}|x^m) \big\| q_\phi(z^{\mathrm{s}}|X) \right) \right].
			\end{aligned}
			\label{eq:reg_enc_static}
		\end{equation}
		Minimizing $\mathcal{R}_{\mathrm{enc}}$ encourages each unimodal encoder $q_{\phi_m}(z^{\mathrm{s}} | x^m)$ to match the fused posterior $q_\phi(z^{\mathrm{s}} | X)$, so that models from different modalities are aligned via a consistent representation of the shared latent variable $z^{\mathrm{s}}$.

	\subsection{Multimodal Framework With Shared-Private Dual Latents}
	\label{subsec:mm_dual}
	
	In \ref{subsec:mm_shared}, we adopt a shared representation as the latent variable for all modalities. Through a specially designed regularization procedure, we theoretically achieve partial decoupling among the training processes of different modalities, thereby supporting the modular use of modality experts. However, this design requires the shared representation $z^{\mathrm{s}}$ to not only capture cross-modal features but also accommodate modality-specific variations. Overemphasizing cross-modal feature learning may prevent the model from extracting sufficiently informative features in unimodal settings, thereby limiting its ability to capture fine-grained details. Conversely, overemphasizing modality-specific characteristics may hinder semantic alignment across modalities.
	
	To eliminate this tension at the structural level, we partition the latent representation into two types in our design. The shared latent $z^{\mathrm{s}}$, as in \ref{subsec:mm_shared}, is used to capture cross-modal features, whereas the private latents $\{z^{\mathrm{p},m}\}_{m\in\mathcal{M}}$ focus on modality-specific attributes. As in \ref{subsec:mm_shared}, we continue to consider the case $T=1$, i.e., we do not model temporal dependencies. Accordingly, the time index $t$ of all variables is fixed to $1$ and is omitted for brevity.
	
	Under this specialization, the joint distribution from \eqref{eq:unified_model} can be simplified as
	\begin{equation}
		\begin{aligned}
			& p_\theta\left(y,X,z^{\mathrm{s}},Z^{\mathrm{p}}\right)\\
			&\quad = p(z^{\mathrm{s}}) p_{\theta_y}(y|z^{\mathrm{s}}) \prod_{m\in\mathcal{M}} p(z^{\mathrm{p},m}) p_{\theta_m}\left(x^m \big| z^{\mathrm{s}}, z^{\mathrm{p},m}\right),
		\end{aligned}
		\label{eq:dual_joint}
	\end{equation}
	where $Z^{\mathrm{p}} \triangleq \{z^{\mathrm{p},m}\}_{m\in\mathcal{M}}$ and $X \triangleq \{x^m\}_{m\in\mathcal{M}}$.
	Compared with \ref{subsec:mm_shared}, each modality likelihood is allowed to depend on an additional private component $z^{\mathrm{p},m}$. This preserves the key modular structure that each modality admits its own decoder conditioned on latent variables, while avoiding the need to force modality-specific effects into $z^{\mathrm{s}}$.
	
	The conditional-independence principle in \eqref{eq:dec_factor_static} is instantiated by the modality-factorized likelihood
	\begin{equation}
		p_\theta\left(X \big| z^{\mathrm{s}}, Z^{\mathrm{p}}\right) = \prod_{m\in\mathcal{M}} p_{\theta_m}\left(x^m \big| z^{\mathrm{s}}, z^{\mathrm{p},m}\right),
		\label{eq:dual_dec_factor}
	\end{equation}
	which implies that cross-modal interactions in the generative path are mediated only through the shared component $z^{\mathrm{s}}$. The private components $z^{\mathrm{p},m}$ affect only their corresponding modalities, thereby isolating modality-specific variability from the shared interface.
	
	\subsection{Sequential Multimodal Scenario With Dual Latents}
	\label{subsec:mm_dual_seq}
	In the preceding analysis, for simplicity, we set the sequence length to $T=1$, i.e., we do not model temporal dependencies. However, in practical RF sensing and ISAC scenarios, many tasks require temporal features. Therefore, in this section, we remove the restriction on the sequence length and introduce the state variables $\{h_t\}_{t=1}^T$ to endow our representation model with the capability to capture temporal dynamics.
	
	Let $X_t \triangleq \{x_t^m\}_{m\in\mathcal{M}}$ and $Z_t^{\mathrm{p}} \triangleq \{z_t^{\mathrm{p},m}\}_{m\in\mathcal{M}}$.
	We consider a recurrent generative process in which the shared latent $z_t^{\mathrm{s}}$ and each private latent $z_t^{\mathrm{p},m}$
	follow history-dependent priors, and each modality is generated conditionally independently given $(z_t^{\mathrm{s}}, z_t^{\mathrm{p},m}, h_{t-1})$.
	Under a first-order Markov assumption, conditioned on $h_{t-1}$, the current-step latent variables and observations are independent of earlier time indices,
	so that all history dependence is mediated through $h_{t-1}$.
	
	We parameterize the state transition as a conditional distribution $p_{\vartheta_h}(h_t | h_{t-1}, z_t^{\mathrm{s}}, Z_t^{\mathrm{p}})$, whose parameters are produced by a recurrent network.
	
	We further factorize the transition term in \eqref{eq:unified_model} as a history-conditioned prior over shared and private latents together with the state transition:
	\begin{equation}
		\begin{aligned}
			& p_{\theta_m,\theta_y,\vartheta}\left(z_t^{\mathrm{s}}, Z_t^{\mathrm{p}}, h_t | h_{t-1}\right)\\
			&\quad \triangleq
			\prod_{m\in\mathcal{M}}
			p_{\vartheta_{p,m}}\left(z_t^{\mathrm{p},m} | h_{t-1}\right) p_{\vartheta_h}\left(h_t | h_{t-1}, z_t^{\mathrm{s}}, Z_t^{\mathrm{p}}\right) \\
			&\times p_{\vartheta_s}\left(z_t^{\mathrm{s}} | h_{t-1}\right).
		\end{aligned}
		\label{eq:seq_trans_factor}
	\end{equation}
	In \eqref{eq:seq_trans_factor}, the history-conditioned priors $p_{\vartheta_s}(z_t^{\mathrm{s}} | h_{t-1})$ and
	$\{p_{\vartheta_{p,m}}(z_t^{\mathrm{p},m} | h_{t-1})\}_{m\in\mathcal{M}}$ are parameterized as lightweight networks that share the same recurrent input $h_{t-1}$.
	
	Consequently, the sequential joint distribution admits
	\begin{equation}
		\begin{aligned}
			&p_\theta\left(y, X_{1:T}, Z_{1:T}^{\mathrm{s}}, Z_{1:T}^{\mathrm{p}}, H_{0:T}\right) \\
			&\;= p(h_0) \prod_{t=1}^{T} p_{\vartheta_s}\left(z_t^{\mathrm{s}} | h_{t-1}\right) \bigg[\\
			&\quad \prod_{m\in\mathcal{M}} p_{\vartheta_{p,m}} \left(z_t^{\mathrm{p},m} | h_{t-1}\right) p_{\vartheta_h}\left(h_t | h_{t-1}, z_t^{\mathrm{s}}, Z_t^{\mathrm{p}}\right) \bigg]\\
			&\;\times \prod_{t=1}^{T}\prod_{m\in\mathcal{M}} p_{\theta_m}\left(x_t^m | z_t^{\mathrm{s}}, z_t^{\mathrm{p},m}, h_{t-1}\right) \cdot p_{\theta_y}\left(y | Z_{1:T}^{\mathrm{s}}, H_{0:T}\right).
		\end{aligned}
		\label{eq:seq_joint}
	\end{equation}
	
	To enable variational inference, we approximate the following unimodal posteriors using neural networks:
	\begin{equation}
		q_{\phi_m}\left(z_t^{\mathrm{s}} | x_t^m, h_{t-1}\right),
		\quad
		q_{\psi_m}\left(z_t^{\mathrm{p},m} | x_t^m, h_{t-1}\right),
		\label{eq:seq_unimodal_posts}
	\end{equation}
	where $\phi_m$ and $\psi_m$ parameterize the shared and private encoders, respectively.
	Following the same rationale as in \ref{subsec:mm_shared} and \ref{subsec:mm_dual}, we aggregate the shared posterior via a PoE:
	\begin{equation}
		\begin{aligned}
			&q_{\phi}\left(z_t^{\mathrm{s}} | X_t, h_{t-1}\right)\\
			&\quad= \frac{1}{\mathcal{Z}_t} p_{\vartheta_s}\left(z_t^{\mathrm{s}} | h_{t-1}\right) \prod_{m\in\mathcal{M}} q_{\phi_m}\left(z_t^{\mathrm{s}} | x_t^m, h_{t-1}\right)^{\alpha_m}.
		\end{aligned}
		\label{eq:seq_poe}
	\end{equation}
	
	We then define the sequential variational posterior by applying the shared and private inference structure of \ref{subsec:mm_shared}-\ref{subsec:mm_dual} at each time step: the shared latent $z_t^{\mathrm{s}}$ is inferred from all available modalities through the fused posterior in \eqref{eq:seq_poe}, while each private latent $z_t^{\mathrm{p},m}$ remains unimodal and is never fused across modalities. To avoid introducing an additional inference network for the recurrent state, we tie the state factor to the generative transition. The resulting variational posterior is
	\begin{equation}
		\begin{aligned}
			&q_{\phi,\psi}\left(Z_{1:T}^{\mathrm{s}}, Z_{1:T}^{\mathrm{p}}, H_{0:T} | X_{1:T}\right) \\
			&\quad= p(h_0) \prod_{t=1}^{T} \Bigg[ q_{\phi}\left(z_t^{\mathrm{s}} | X_t, h_{t-1}\right)\\
			&\qquad \times  \prod_{m\in\mathcal{M}}	q_{\psi_m}\left(z_t^{\mathrm{p},m} | x_t^m, h_{t-1}\right) p_{\vartheta_h}\left(h_t | h_{t-1}, z_t^{\mathrm{s}}, Z_t^{\mathrm{p}}\right) \Bigg].
		\end{aligned}
		\label{eq:seq_q}
	\end{equation}

    \subsection{Objective Function and Problem Optimization}
	\label{subsec:object}
	
	Consistent with \ref{subsec:gen_vi}, the learning objective is to maximize the ELBO.
	Substituting \eqref{eq:seq_joint} and \eqref{eq:seq_q} into the unified ELBO in \eqref{eq:unified_elbo}, we obtain the sequential objective
	\begin{equation}
		\begin{aligned}
			&\mathcal{L}_{\mathrm{seq}}(\theta,\phi,\psi) = \mathbb{E}_{q_{\phi,\psi}} \Bigg[ \log p_{\theta_y}\left(y | Z_{1:T}^{\mathrm{s}}, H_{0:T}\right)\\
			& + \sum_{t=1}^{T}\sum_{m\in\mathcal{M}} \log p_{\theta_m}\left(x_t^m | z_t^{\mathrm{s}}, z_t^{\mathrm{p},m}, h_{t-1}\right) \Bigg] \\
			&- \sum_{t=1}^{T} D_{\mathrm{KL}}\left( q_{\phi}\left(z_t^{\mathrm{s}} | X_t, h_{t-1}\right) \big\| p_{\vartheta_s}\left(z_t^{\mathrm{s}} | h_{t-1}\right) \right) \\
			&- \sum_{t=1}^{T}\sum_{m\in\mathcal{M}} D_{\mathrm{KL}}\left( q_{\psi_m}\left(z_t^{\mathrm{p},m} | x_t^m, h_{t-1}\right) \big\| p_{\vartheta_{p,m}}\left(z_t^{\mathrm{p},m} | h_{t-1}\right)\right),
		\end{aligned}
		\label{eq:seq_elbo}
	\end{equation}
	where the expectation is taken with respect to $q_{\phi,\psi}(Z_{1:T}^{\mathrm{s}}, Z_{1:T}^{\mathrm{p}}, H_{0:T} | X_{1:T})$ in \eqref{eq:seq_q}.
	To incorporate the regularization in \eqref{eq:reg_dec_static} and \eqref{eq:reg_enc_static}, we consider the following criterion:
	\begin{equation}
		\max_{\theta,\phi}\quad \mathcal{L}_{\mathrm{seq}}(\theta,\phi) - \lambda_{\mathrm{enc}}\mathcal{R}_{\mathrm{enc}}(\phi) - \lambda_{\mathrm{dec}}\mathcal{R}_{\mathrm{dec}}(\theta),
		\label{eq:mm_obj}
	\end{equation}
	where $\lambda_{\mathrm{enc}}\ge 0, \lambda_{\mathrm{dec}}\ge 0$ control the strength of regularization.

    \begin{remark}
	With our design, the optimization objective in \eqref{eq:mm_obj} is explicitly decomposed into two parts. The first part is the ELBO, which can be decomposed by modality-specific terms related to unimodal data and requires no cross-modality interaction. The second part consists of encoder and decoder regularization terms, which rely on semantically aligned multimodal data. Inspired by the warm-up strategy and dynamic weighting trick in deep learning, we set $\lambda_{\mathrm{enc}}$ and $\lambda_{\mathrm{dec}}$ to $0$ at the beginning of training, allowing the model to focus on learning modality-specific representational capacity. As training progresses, $\lambda_{\mathrm{enc}}$ and $\lambda_{\mathrm{dec}}$ are increased to activate regularization and impose the desired constraints. The schedules of $\lambda_{\mathrm{enc}}$ and $\lambda_{\mathrm{dec}}$ can be either continuous with respect to training epochs or stepwised.
    \end{remark}
	
	    	\begin{algorithm}[t!]
		\caption{Unimodal Pretraining for Modality $m$}
		\label{alg:uni_pretrain}
		\KwIn{Mini-batch $\{(x_{1:T}^{m},y)\}$; learning rate $\eta$.}
		\KwOut{$\theta_y,\theta_m,\phi_m,\psi_m,\vartheta_{p,m},\vartheta_{s,m},\vartheta_{h,m}$.}
		\While{not converged}{
			Sample a mini-batch $\{(x_{1:T}^{m},y)\}$\;
			Set $\mathcal{J}\leftarrow 0$\;
			\ForEach{$(x_{1:T}^{m},y)$ in the mini-batch}{
				\For{$t\leftarrow 1$ \KwTo $T$}{
					Compute $q_{\phi_m}(z_t^{\mathrm{s}}|x_t^{m},h_{t-1})$ and
					$q_{\psi_m}(z_t^{\mathrm{p},m}|x_t^{m},h_{t-1})$\;
					Compute $p_{\vartheta_{s,m}}(z_t^{\mathrm{s}}|h_{t-1})$,
					$p_{\vartheta_{p,m}}(z_t^{\mathrm{p},m}|h_{t-1})$,
					$p_{\vartheta_{h,m}}(h_t|h_{t-1},z_t^{\mathrm{s}},z_t^{\mathrm{p},m})$\;
					Sample $z_t^{\mathrm{s}} \sim q_{\phi_m}(z_t^{\mathrm{s}}|x_t^{m},h_{t-1})$\;
					Sample $z_t^{\mathrm{p},m} \sim q_{\psi_m}(z_t^{\mathrm{p},m}|x_t^{m},h_{t-1})$\;
					Sample $h_t \sim p_{\vartheta_{h,m}}(h_t|h_{t-1},z_t^{\mathrm{s}},z_t^{\mathrm{p},m})$\;
					Update $\widehat{\mathcal{L}}$ by restricting \eqref{eq:seq_elbo} to modality $m$ and using the unimodal shared posterior in \eqref{eq:seq_unimodal_posts}\;
				}
				Update $\widehat{\mathcal{L}}$ by adding the task term in \eqref{eq:seq_elbo}\;
				$\mathcal{J}\leftarrow \mathcal{J}-\widehat{\mathcal{L}}$\;
			}
			$(\theta_y,\theta_m,\phi_m,\psi_m,\vartheta_{p,m},\vartheta_{s,m},\vartheta_{h,m})
			\leftarrow
			(\theta_y,\theta_m,\phi_m,\psi_m,\vartheta_{p,m},\vartheta_{s,m},\vartheta_{h,m})
			- \eta \nabla \mathcal{J}$\;
		}
	\end{algorithm}

	\begin{algorithm}[t!]
		\caption{Multimodal Fine-Tuning With PoE Fusion and Interface Regularization}
		\label{alg:mm_finetune}
		\KwIn{Mini-batch $\{(X_{1:T}^{(i)},y^{(i)})\}_{i=1}^{B}$; weights $\{\alpha_m\}_{m\in\mathcal{M}}$; $\lambda_{\mathrm{enc}},\lambda_{\mathrm{dec}}$; learning rate $\eta$.}
		\KwOut{$\theta_y,\vartheta_h,\vartheta_s,\{\theta_m,\phi_m,\psi_m,\vartheta_{p,m}\}_{m\in\mathcal{M}}$.}
		Initialize $\{\theta_m,\phi_m,\psi_m,\vartheta_{p,m}\}_{m\in\mathcal{M}}$ from unimodal pretrained models\;
		Initialize $\vartheta_h \leftarrow \mathrm{Avg}(\{\vartheta_{h,m}\}_{m\in\mathcal{M}})$ and
		$\vartheta_s \leftarrow \mathrm{Avg}(\{\vartheta_{s,m}\}_{m\in\mathcal{M}})$\;
		Initialize $\theta_y$\;
		\While{not converged}{
			Sample a mini-batch $\{(X_{1:T}^{(i)},y^{(i)})\}_{i=1}^{B}$\;
			Set $\mathcal{J}\leftarrow 0$\;
			\For{$i\leftarrow 1$ \KwTo $B$}{
				\For{$t\leftarrow 1$ \KwTo $T$}{
					Define $\mathcal{M}_t^{(i)} \subseteq \mathcal{M}$\;
					\ForEach{$m\in\mathcal{M}_t^{(i)}$}{
						Compute $q_{\phi_m}(z_t^{\mathrm{s}}|x_t^{m,(i)},h_{t-1}^{(i)})$ and
						$q_{\psi_m}(z_t^{\mathrm{p},m}|x_t^{m,(i)},h_{t-1}^{(i)})$ by \eqref{eq:seq_unimodal_posts}\;
					}
					Compute $q_{\phi}(z_t^{\mathrm{s}}|X_t^{(i)},h_{t-1}^{(i)})$ by \eqref{eq:seq_poe} using $m\in\mathcal{M}_t^{(i)}$\;
					Sample $z_t^{\mathrm{s},(i)}$ and $\{z_t^{\mathrm{p},m,(i)}\}_{m\in\mathcal{M}_t^{(i)}}$\;
					Sample $h_t^{(i)}$ by the state factor in \eqref{eq:seq_q}\;
					Accumulate $\widehat{\mathcal{L}}^{(i)}$ by \eqref{eq:seq_elbo} across $m\in\mathcal{M}_t^{(i)}$\;
					Accumulate $\widehat{\mathcal{R}}_{\mathrm{enc}}^{(i)}$ by \eqref{eq:reg_enc_static}\;
					Update residual features by \eqref{eq:dec_residual_feat} and
					accumulate $\widehat{\mathcal{R}}_{\mathrm{dec}}^{(i)}$ by \eqref{eq:reg_dec_static}\;
				}
				$\mathcal{J}\leftarrow \mathcal{J}
				-\widehat{\mathcal{L}}^{(i)}
				+\lambda_{\mathrm{enc}}\widehat{\mathcal{R}}_{\mathrm{enc}}^{(i)}
				+\lambda_{\mathrm{dec}}\widehat{\mathcal{R}}_{\mathrm{dec}}^{(i)}$\;
			}
			$(\theta_y,\vartheta_h,\vartheta_s,\{\theta_m,\phi_m,\psi_m,\vartheta_{p,m}\})
			\leftarrow
			(\theta_y,\vartheta_h,\vartheta_s,\{\theta_m,\phi_m,\psi_m,\vartheta_{p,m}\})
			- \eta \nabla \mathcal{J}$\;
		}
	\end{algorithm}
	
	When $\lambda_{\mathrm{enc}}$ and $\lambda_{\mathrm{dec}}$ follow a stepwise schedule, we obtain an explicit separation between unimodal feature learning and multimodal alignment. When $\lambda_{\mathrm{enc}}$ and $\lambda_{\mathrm{dec}}$ keep $0$, training does not depend on any semantically aligned multimodal data. Accordingly, we divide the training procedure into two stages:
    \begin{itemize}
	\item{\textbf{Stage I: Unimodal ELBO Pretraining}:}
	For each modality $m$, we maximize a unimodal ELBO using unimodal samples. Stage I trains each modality expert independently, thereby reducing the dependence on paired multimodal observations.
	
	\item{\textbf{Stage II: Multimodal ELBO Fine-tuning With Posterior Alignment}:}
	Starting from the pretrained experts, we maximize \eqref{eq:mm_obj} using aligned multimodal samples.
	In this stage, the fused posterior in \eqref{eq:mm_poe} pools multimodal evidence, and $\mathcal{R}_{\mathrm{enc}}(\phi)$ and $\mathcal{R}_{\mathrm{dec}}(\theta)$ in \eqref{eq:reg_enc_static} and \eqref{eq:reg_dec_static} enforces modality experts decoupled while remains compatible with multimodal fusion. During this process, knowledge transfer across modalities is achieved: a modality expert can benefit from the others, and the resulting PoE representation becomes closer to the true multimodal joint posterior.
	\end{itemize}
    
    As a conclusion, the detailed processes for unimodal pretraining and multimodal fine-tuning are summarized in Algorithms \ref{alg:uni_pretrain} and \ref{alg:mm_finetune}, respectively.
	
	\section{Model Design}
    \label{sec:model_design}

	In this section, we further specify the designs of the unimodal sub-module models and the task model. For the unimodal sub-module models, we propose a network architecture for the RF sensing scenario. For the task model, we provide a general network design for classification problems.
    
	\subsection{Modality Model}
	\label{subsec:mod_exp}
	
	Our algorithmic framework provides a modular fusion approach for multimodal sensing. Specifically, we only need to design, for each modality, an encoder and decoder for our proposed VAE-based framework that can operate in a unimodal setting. After unimodal pretraining, the model will be fine-tuned using a small amount of multimodal data. Given that RF signals constitute the most critical modality in ISAC scenarios, we present a detailed design example of the RF-modality sub-model under the proposed framework.
	
	
	The RF signal of each frame can be expressed as a complex-valued tensor with three dimensions, whose indices correspond to the receive-antenna index, the sample index within each chirp, and the chirp index within each frame, respectively. In practical deployments, variations in channel conditions and spectrum occupancy may cause the BS to operate with different scanning configurations. To accommodate such variability, we first apply zero padding along the second dimension of the data cube $\mathbf B$ to obtain $\mathbf B'$, thereby ensuring a fixed input size for the network in subsequent processing.
	
	The data cube $\mathbf B$ and its padded version $\mathbf B'$ exhibit variations along the sample and chirp dimensions that characterize the dynamic behaviors in fast time and slow time, respectively. We denote the padded length of $\mathbf B'$ along the chirp dimension by $C'$, and the length along the sample dimension by $S$. Owing to the physical properties of the RF sensing measurement mechanism, applying the discrete Fourier transform (DFT) along the corresponding dimensions in signal processing can explicitly decouple the information across different axes. Specifically, the frequency components along the fast-time dimension are approximately linear with respect to the echo delay, thereby yielding a range representation; the phase shifts across slow time reflect Doppler-induced frequency offsets, which correspond to radial velocity.
	
	Motivated by this principle, we design our network architecture based on the DFT. We define a window matrix
	\begin{equation}
		\mathbf W = \mathbf w_s \mathbf w_c^\mathrm H,
	\end{equation}
	where $\mathbf w_s$ is an $S$-dimensional column vector and $\mathbf w_c$ is a $C'$-dimensional column vector. Accordingly, the windowed data cube can be written as
	\begin{equation}
		\mathbf B'_w = \mathbf B' \odot \mathbf W,
	\end{equation}
	where $\odot$ represents the Hadamard product.
	
	We enhance the role of the window function by introducing a learnable residual term in the log domain. Specifically, we let the windowing vectors $\mathbf w_s$ and $\mathbf w_c$ satisfy:
	\begin{equation}
		\begin{aligned}
			\mathbf w_s &= \frac{\mathbf w_s^o \odot \exp(\mathbf \delta_s)}{\left\|\mathbf w_s^o \odot \exp(\mathbf \delta_s) \right\|_2}\\
			\mathbf w_c &= \frac{\mathbf w_c^o \odot \exp(\mathbf \delta_c)}{\left\|\mathbf w_c^o \odot \exp(\mathbf \delta_c) \right\|_2},
		\end{aligned}
	\end{equation}
	where $\mathbf w_s^o$ and $\mathbf w_c^o$ are unlearnable windowing vectors designed with commonly used window functions like Hamming window or Hann window, and $\mathbf \delta_s, \mathbf \delta_c$ are learnable parameters.
	
	We define $\mathbf X_m$ as the baseband data received by the $m$-th antenna with size $S\times C'$. We then compute
	\begin{equation}
		\mathbf Y_m = \mathbf F_r \mathbf X_m \mathbf F_d^{\mathrm H},
	\end{equation}
	where $\mathbf F_r \in \mathbb C^{S\times S}$ and $\mathbf F_d \in \mathbb C^{C'\times C'}$. We further define $\mathbf F_r$ and $\mathbf F_d$ to satisfy
	\begin{equation}
		\begin{aligned}
			\mathbf F_r &= \mathbf F_r^o + \Delta_r, \quad \mathbf F_r^\mathrm H \mathbf F_r \approx \mathbf I\\
			\mathbf F_d &= \mathbf F_d^o + \Delta_d, \quad \mathbf F_d^\mathrm H \mathbf F_d \approx \mathbf I,
		\end{aligned}
	\end{equation}
	where $\mathbf F_r^o$ and $\mathbf F_d^o$ are Fourier matrices, $\mathbf \Delta_r$ and $\mathbf \Delta_d$ are learnable complex parameters. Considering that, in our framework, the encoder extracts features to obtain latent representations, while the decoder must be able to reconstruct the original data, and the predictor must retrieve sufficient information from the latent space, we further require that the matrices $\mathbf F_r$ and $\mathbf F_d$ be approximately unitary, so as to prevent singularities during data processing.
	
	After the learnable windowing and the DFT, we obtain $N_{rx}$ Range-Doppler domain data matrices $\mathbf Y_m$. Since $\mathbf Y_m$ is complex-valued, we separate it into its real and imaginary parts and treat them as two feature channels. We then concatenate these channels along the first dimension associated with the receive antennas, yielding a real-valued tensor $\mathbf Z \in \mathbb{R}^{2N_{rx} \times S \times C'}$.
	
	We adapt ResNet18\cite{ResNet} for feature extraction from the radar data block $\mathbf Z$. Since $\mathbf Z$ has $2N_{rx}$ channels, we modify the input layer of ResNet18 to accept $2N_{rx}$ input channels. In addition, unlike natural images, radar data blocks contain a larger number of fine-grained structural patterns; therefore, we adopt smaller kernel sizes and stride values. We also replace Batch Normalization with Group Normalization to better exploit the benefits of larger channel numbers.
	
	The decoder is designed as a mirror of the encoder. Specifically, the ResNet18 used for feature extraction in the encoder is reconfigured with its input-output mapping reversed to reconstruct the original data. In this process, pooling operations are replaced with PixelShuffle\cite{PixelShuffle} to increase the spatial resolution. The learnable DFT is replaced with a learnable IDFT, and windowing is replaced with de-windowing.
	
	\subsection{Task Model}
	
	By using the encoder and decoder networks in \ref{subsec:mod_exp} to approximate the posterior and likelihood terms in Algorithm \ref{alg:uni_pretrain}, we obtain a feature representation network that maps any modality into a unified representation space. Using Algorithm \ref{alg:mm_finetune}, we further fine-tune this network to achieve semantic alignment across modalities, thereby producing an aggregated multimodal representation. However, in practice, our ultimate objective is often downstream tasks, such as classification and prediction, rather than representation learning alone. Therefore, we additionally require a task model that converts the multimodal sequential representations into the final task outputs. However, since the latent variables already serve as representations that capture high-level features, in most cases only a lightweight network is needed, allowing our representation model to function as a feature extractor for a variety of downstream tasks.
	
	At each time step $t$, the multimodal inference module constructs a fused posterior for the shared
	latent variable via PoE fusion according to \eqref{eq:seq_poe}. Since the shared latent variable is stochastic, we use its posterior expectation as the input to the task model:
	\begin{equation}
		\tilde{z}^{\mathrm s}_t = \mathbb{E}_{q_{\phi}} \left[z^{\mathrm s}_t | X_t, h_{t-1}\right].
		\label{eq:task_mean}
	\end{equation}
	
	To capture temporal dependencies, the task model needs an aggregated representation over time derived from the time series of the shared latent representations and the state variables:
	\begin{equation}
		c_t = \mathrm{Agg}\left(\{\tilde{z}^{\mathrm s}_k\}_{k=t-N}^{t}, h_{t-1}\right),
		\label{eq:task_agg}
	\end{equation}
	where $N$ denotes the observation window size of the task model. The operator $\mathrm{Agg}(\cdot)$ aggregates multiple input feature vectors into a single representation. In our design, we adopt the simplest choice, where $\mathrm{Agg}(\cdot)$ performs a one-dimensional concatenation of input vectors. For more complex tasks, $\mathrm{Agg}(\cdot)$ can be instantiated using other architectures, such as an MLP, one-dimensional convolution, or a Transformer.
	
	Given the aggregated representation $c_t$, the task model constructs a predictive distribution over the outputs as:
	\begin{equation}
		p_{\theta_y}\left(y | Z_{1:t}^{\mathrm s}, H_{0:t-1}\right) = f_{\theta_y}(c_t),
		\label{eq:task_output}
	\end{equation}
	where $f_{\theta_y}(\cdot)$ denotes a neural network parameterized by $\theta_y$ to approximate the ground-truth distribution. In our design, we instantiate $f_{\theta_y}(\cdot)$ as a four-layer MLP with GELU activations, whose hidden-layer widths are $512$, $1024$, $256$, and $64$, respectively.
	
	It is worth noting that the task model is not trained jointly with the representation model. Instead, after the representation model is trained and frozen, the task model is trained separately. In this way, we use the representation model as a general-purpose feature extractor, thereby decoupling upstream feature extraction from downstream task-specific learning. This separation allows the representation model to be developed prior to downstream tasks and to leverage abundant and low cost unlabeled data collected during system operation. Although our algorithmic is designed to support representation learning with task ground truth, we can still train the model without tags by treating $y$ as a task-irrelevant random variable.
	
	Benefiting from the properties of the PoE fusion used in \eqref{eq:seq_poe}, both the representation model and the task model can effectively handle missing-modality scenarios without any architectural modification. When a modality is unavailable, the task-model inference procedure described in \eqref{eq:task_mean}-\eqref{eq:task_output} remains unchanged. In this case, the representation model aggregates fewer distributions to form $q_{\phi} \left(z_t^{\mathrm{s}} | X_t,h_{t-1}\right)$, while interacting with the task model through the same data interface.
	
	\section{Numerical Experiments}
    \label{sec:experiments}
	
	To evaluate the performance of our framework in realistic environments, we conduct experiments using the DeepSense6G\cite{DeepSense6G} dataset. This dataset contains multiple data collections captured in real-world scenarios and supports the evaluation of various task types. In our study, we focus on the multimodal beam prediction task, corresponding to Scenes 31, 32, 33, and 34 of the dataset. These scenes were recorded on a two-lane bidirectional urban road, where Scenes 31 and 32 were collected during daytime, and Scenes 33 and 34 were collected at night.
	
	Each scene contains a mobile unit mounted on a vehicle and a stationary unit deployed at a BS, along with other objects commonly present in urban street environments, such as surrounding vehicles and pedestrians. The vehicle unit is equipped with an omnidirectional millimeter-wave transmitter to enable AoA measurements at the base-station receiver, as well as a GPS receiver for obtaining vehicle position information. The base-station unit is equipped with a richer set of sensors, providing multimodal data that include millimeter-wave RF sensor data cubes, RGB images, GPS coordinates and 3D LiDAR point clouds.
	
	Our objective is to achieve multimodal sensing-aided beam prediction by leveraging measurements of the target vehicle obtained from the multimodal sensor suite deployed at the BS. For each sampling instant $t$, we aggregate data from five consecutive time steps, spanning from $t-4$ to $t$, to form a temporal sequence that enables the extraction of the target’s motion dynamics. For the RF signals, RGB images and LiDAR point clouds, observations at all five time steps are available. However, for the GPS-based position measurements, only the data from time steps $t-4$ and $t-3$ are accessible, reflecting the intrinsic limitations in its update rate and reliability. Our multimodal algorithm utilizes these heterogeneous measurements to generate the beamforming strategy at time $t$. Specifically, our algorithm selects the optimal beam index from a predefined 64-beam codebook.
	
	To ensure consistency with other algorithms evaluated on the DeepSense6G dataset, we adopt the distance-based accuracy (DBA) metric \cite{DBA} as our performance indicator. This metric, recommended by the dataset creators, is defined as
	\begin{equation}
		\mathrm{DBA-Score} = \frac{1}{3} (Y_1 + Y_2 + Y_3),
	\end{equation}
	where $Y_i$ denotes the accuracy level corresponding to the $i$-th distance threshold and is defined as
	\begin{equation}
		Y_i = \frac{1}{N}\sum_{n=1}^{N} \mathbf{1}\left( d(\hat{b}_n, b_n) \leq d_i \right),
	\end{equation}
	where $N$ is the total number of beam prediction instances, $\hat{b}_n$ is the predicted beam index for sample $n$, $b_n$ is the corresponding ground-truth beam index, $d(\hat{b}_n, b_n)$ denotes the distance between the predicted and ground-truth beams, $d_i$ is the $i$-th predefined distance threshold, and $\mathbf{1}(\cdot)$ is the indicator function.
	
	\subsection{Performance Evaluation}
	
	\begin{table}[t]
		\centering
		\caption{Comparison of the Proposed Method with Benchmarks Based on DBA-Score}
		\label{tab:dba_comparison}
		\begin{tabular}{l|ccccc}
			\hline
			Method			& S31		& S32		& S33		& S34		& Overall \\
			\hline
			TII\cite{TransformerFusion-Comm}				& 0.7298	& 0.7852	& 0.8462	& 0.8430	& 0.7844 \\
			Avatar\cite{DBA}			& 0.6536	& 0.7074	& 0.8576	& 0.7120	& 0.7162 \\
			BeamTransFuser\cite{BeamTransFuser}	& \textbf{1.0000}	& \textbf{0.9038}	& 0.8988	& 0.8945	& \textbf{0.9129} \\
			QTNs\cite{QTN}			& 0.7605	& 0.8707	& 0.8864	& 0.9124	& - \\
			\hline
			\textbf{Ours (data-efficient)}	& 0.8541	& 0.8765	& 0.8972	& 0.8609	& 0.8762 \\
			\textbf{Ours (data-sufficient)}	& 0.8977	& 0.8620	& \textbf{0.9218}	& \textbf{0.9313}	& 0.8976 \\
			\hline
		\end{tabular}
	\end{table}
	
	We adopt four recent studies for the multimodal beam prediction scenario as benchmarks, namely TII\cite{TransformerFusion-Comm}, Avatar\cite{DBA}, BeamTransFuser\cite{BeamTransFuser} and QTNs\cite{QTN}. We reserve 10\% of the dataset as the test set and use the remaining data for training. For the benchmark methods, the entire training set is used as semantically aligned multimodal data. In contrast, in our method, benefiting from inter-modality decoupling and modularity, we can first pretrain on large-scale unimodal unlabeled data accumulated during system operation, then perform modality-alignment fine-tuning using a small amount of multimodal data, and finally train the task model in a supervised manner using a small amount of labeled data. To emulate this practical workflow, we design two training strategies:
	\begin{enumerate}
		\item \emph{Data-efficient Training:} We randomly split the dataset into two subsets containing $80\%$ and $20\%$ of the samples, respectively. The $80\%$ subset is further separated into unimodal datasets, where samples are no longer jointly aligned across modalities. The remaining $20\%$ subset preserves the original modality alignment of the DeepSense6G dataset. We use the $80\%$ subset for unimodal pretraining and the $20\%$ subset for multimodal alignment.
		\item \emph{Data-sufficient Training:} To assess the upper-bound performance of our method, we fully exploit all available data in this strategy. Specifically, both pretraining and fine-tuning are performed on the complete training dataset.
	\end{enumerate}
	Table \ref{tab:dba_comparison} compares the performance of our algorithmic framework against benchmark methods. When using only $20\%$ of the modality-aligned data, our method achieves performance comparable to the benchmark methods that are trained with the full dataset. When all data are available, our method outperforms the benchmarks in the two nighttime scenarios, S33 and S34.
	
	\begin{table}[t]
		\centering
		\caption{Performance degradation of the Proposed Method with Missing Modalities Based on DBA-Score}
		\label{tab:miss_modal}
		\begin{tabular}{l|ccccc}
			\hline
			Missing Modal	& S31		& S32		& S33		& S34		& Overall \\
			\hline
			RF				& 0.1854	& 0.3043	& 0.4164	& 0.3269	& 0.3224 \\
			GPS				& 0.0916	& 0.1173	& 0.1608	& 0.1328	& 0.1470 \\
			Camera			& 0.0780	& 0.1581	& 0.2036	& 0.0875	& 0.1312 \\
			LiDAR			& 0.0413	& 0.0811	& 0.1309	& 0.1281	& 0.1153 \\
			LiDAR \& Camera	& 0.3173	& 0.4977	& 0.3641	& 0.3723	& 0.4363 \\
			\hline
		\end{tabular}
	\end{table}
	
	We further evaluate the robustness of the proposed method under missing-modality conditions. Using the data-efficient training regime as the baseline, we mask out specific modalities at inference time and measure the resulting DBA scores. The performance degradations relative to the complete-modality setting are summarized in Table \ref{tab:miss_modal}. A larger degradation indicates a greater loss in accuracy.
	
	As shown in Table \ref{tab:miss_modal}, the performance degradation is relatively small when either the camera or LiDAR modality is missing. This is because camera and LiDAR provide highly consistent information, and the missing modality can be partially compensated by the other via cross-modality complementarity. In contrast, the absence of the RF modality leads to a substantially larger degradation, since the information provided by RF cannot be fully complemented by the other modalities, resulting in information loss. Similarly, when both camera and LiDAR are missing, RF alone cannot sufficiently compensate for these two modalities, which also causes a considerable performance drop. Meanwhile, the results in Table \ref{tab:miss_modal} also indicate that, when sufficient data are available to enable modality complementarity, our framework is robust to missing-modality scenarios. Owing to the inherent properties of the PoE architecture, modality absence does not alter the operating procedure or the underlying principle of our framework, thereby preventing the model from behaving in unintended ways.
	
	\subsection{Ablation Study}

    \begin{figure}[t!]
		\centering
		\includegraphics[width=0.8\linewidth]{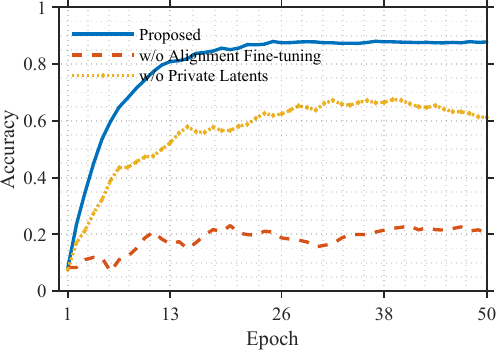}
		\caption{Impact of alignment fine-tuning and dual latents}
		\label{fig:accuracyvsepoch}
	\end{figure}
	
	To investigate the roles of multimodal fine-tuning and the private--shared latent structure, we compare three different settings in the experiment shown in Fig. \ref{fig:accuracyvsepoch}. The figure reports the accuracy evolution of the task model over 50 epochs of supervised training, where the multimodal framework serves as the upstream feature extractor. Compared with the full framework, which converges to an accuracy of 86\%, the baseline model that directly applies PoE aggregation on the latents without modality-alignment fine-tuning suffers from severe convergence difficulty and achieves only about 20\% accuracy. This indicates that unaligned latents cannot be reasonably aggregated via PoE to form effective multimodal representations. In another comparison, the aligned model that only includes the shared latent, but excludes the private latent, reaches only about 60\% accuracy. This suggests that, although the private latent does not participate in multimodal aggregation, it plays an important role in facilitating the shared latent to extract modality-consistent information.
	
\begin{figure}[t!]
		\centering
		\includegraphics[width=0.8\linewidth]{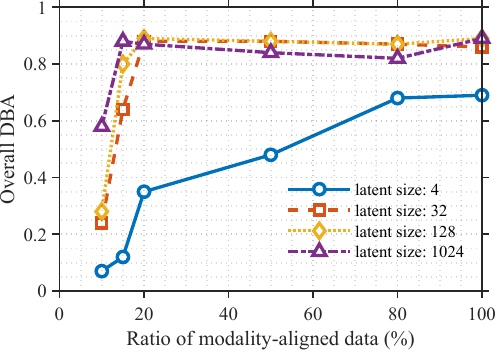}
		\caption{Overall DBA vs. multimodal data ratio.}
		\label{fig:overall_dba_ratio}
	\end{figure}

	To further investigate the effectiveness of the proposed inter-modality decoupling, we evaluate model performance under different proportions of modality-aligned multimodal data and summarize the results in Fig. \ref{fig:overall_dba_ratio}. In these experiments, we use the entire training set for pretraining, and randomly sample different fractions of the data as modality-aligned multimodal samples for alignment fine-tuning. We consider four configurations that span from relatively small to relatively large latent dimensionalities.
	
	Figure \ref{fig:overall_dba_ratio} shows that, with the total amount of training data fixed, allocating a larger portion of data to fine-tuning generally improves the final performance. However, when the latent dimensionality is sufficiently large, the benefit of additional multimodal data exhibits diminishing returns. Specifically, when the latent size exceeds $32$, using more than $20\%$ multimodal aligned data does not yield a noticeable performance gain. For overly small latent sizes, the model performance drops significantly due to limited representational capacity of the latent space, and the diminishing-return regime appears later as more multimodal data are added. Conversely, when the latent size is excessively large, using too much multimodal data may lead to a slight performance degradation. This may be because an overly large latent space makes the latent variables underdetermined, which can introduce training instability.
	
\begin{figure}[t]
		\centering
		\includegraphics[width=0.8\linewidth]{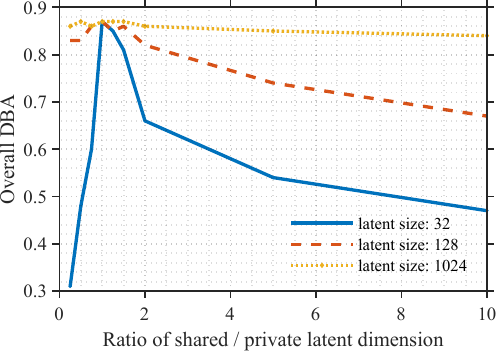}
		\caption{Overall DBA vs. Shared / Private latent dimension ratio.}
		\label{fig:overall_dba_latent_ratio}
	\end{figure}

	In another set of experiments, we examine how the ratio between the shared latent and the private latent affects performance when the total latent dimensionality is kept constant. The results are shown in Fig. \ref{fig:overall_dba_latent_ratio}. When the total latent size is small, the performance reaches its maximum when the two latent components have nearly equal dimensionalities. As the total latent size increases, this peak becomes progressively less pronounced. This is because, with a limited total latent budget, ratios that deviate substantially from 1 allocate very few dimensions to one of the latent components, which then becomes the bottleneck of the overall model. When the total latent dimensionality is sufficiently large, both latent components have redundant expressive capacity, making the performance less sensitive to their relative sizes.
	
	The results in Figs. \ref{fig:overall_dba_ratio} and \ref{fig:overall_dba_latent_ratio} suggest using a slightly over-parameterized latent space to provide sufficient representational capacity, while avoiding excessively large dimensionalities to reduce computational overhead and to mitigate potential divergence caused by underdetermined latent variables. Moreover, provided that the model is sufficiently pretrained in unimodal settings, a moderate amount of modality-aligned multimodal data is adequate to reach near-optimal performance, and additional multimodal data yields no substantial improvement.
	
	\section{Conclusion}
    \label{sec:conclusion}
    
	This paper proposed a variational-inference-based modular multimodal ISAC framework that introduces a shared latent interface to decouple modality-specific representation learning from cross-modal semantic alignment. Multimodal evidence is pooled via a PoE shared posterior, while posterior-consistency and decoder-side decorrelation regularizers are used to improve latent compatibility and suppress residual cross-modal dependence, enabling a two-stage training workflow with inexpensive unimodal pretraining and data-efficient multimodal fine-tuning. Experiments on the DeepSense6G dataset showed that the proposed framework achieved performance comparable to recent benchmarks while using only $20\%$ of the multimodal data, and it outperformed all benchmarks in two scenarios when the full multimodal dataset was available.

	\bibliographystyle{IEEEtran}
	\bibliography{reference.bib}

\end{document}